\documentstyle[11pt]{article}
\bibliography{plain}
\title{\bf Hierarchy of random deterministic chaotic maps with an
 invariant measure}
\vspace{20mm}
\author{M. A. Jafarizadeh$^{a,b,c}$\thanks{E-mail:jafarzadeh@ark.tabrizu.ac.ir}
, S.Behnia$^{b,e}$ .\\ $^a${\small Department of Theoretical
Physics and Astrophysics, Tabriz University, Tabriz 51664, Iran.}
\\ $^b${\small Institute for Studies in Theoretical Physics and
Mathematics, Teheran 19395-1795, Iran.} \\ $^c${\small Pure and
Applied Science Research Center, Tabriz 51664, Iran.} \\  $^e $
{\small Department of Physics, IAU, Urmia, Iran.}}
\begin{document}
\maketitle
\begin{abstract}
Hierarchy of one and many-parameter families of random
trigonometric chaotic maps and one-parameter random elliptic
chaotic maps of $\bf{cn}$ type with an invariant measure have been
introduced. Using the invariant measure (Sinai-Ruelle-Bowen
measure), the Kolmogrov-Sinai entropy of the random chaotic maps
have been calculated analytically, where the numerical simulations
support the results \noindent .
 {\bf PACs numbers:05.45.Ra, 05.45.Jn, 05.45.Tp }
\end{abstract}
\pagebreak \vspace{7cm}
\section{Introduction}
The problem of the transition to chaos in deterministic systems
has been the subject of much interest, and, for low-dimensional
dynamics, it has been found that this transition most often occurs
via a small number of often observed routes (e.g., period doubling
and intermittency). Usually the analytic calculation of invariant
measure of dynamical systems is a nontrivial task, hence there are
limited number of  maps  with invariant measure like, Ulam-Von
Neumann map \cite{Ulam}, chebyshev maps \cite{Adler},
Katsura-Fukuda map \cite{Kasta}, piecewise parabolic map
\cite{Gyo}, Tent map\cite{dorf1}, Elliptic map \cite{umeno1} and
finally  hierarchy of one and many-parameter families of random
trigonometric chaotic and one-parameter random elliptic chaotic
maps of $\bf{cn}$ type and their coupling  \cite{jafar1, jafar2,
jafar3, jafar4}.\\ Here in this work we give a new hierarchy of
random chaotic maps with an invariant measure, where using the
invariant measure we discuss analytically the transition to chaos
in these random dynamical systems. Random maps have attracted the
attention of physicists in the realm of theoretical biology,
disordered systems, and cellular automata for its possible
application to studies of DNA replication, cell differentiation,
and evolution theory \cite{coste, Derrida}. Additionally, random
maps are of interest as models of convection by temporarily
irregular fluid flows \cite{ott}.\\ In this paper we consider
random maps where, on each iteration, the map function
$\Phi(x,\alpha)$ is chosen randomly from a hierarchy of chaotic
maps with an invariant measure which are introduced in our
previous papers \cite{jafar1, jafar2, jafar4}.\\ There are two
noticeable advantages of random chaotic maps that are presented
through this article. First they are measureable  dynamical system
so they can be studied analytically. Second, they have the
property of being either chaotic or having stable period one fixed
point.
 \\ The paper is organized as follows. Section 2 is devoted to
introduction of the random map models. Then, in the section 3, we
introduce Sinai-Ruelle-Bowen (SRB) measure of random chaotic maps.
The calculation of Kolmogorov-Sinai (KS) entropy  of random
chaotic maps via their SRB-measure is presented in section 4.
Section 5 is devoted to the Lyapunov characteristic exponent. The
paper is ended with a brief conclusion in section 6.
\section{Hierarchy of one and many-parameter random chaotic maps of
trigonometric and elliptic types } The random chaotic map can be
obtained  via random choice from an ensemble of chaotic maps
according to some probability distribution. Therefore for a given
ensemble of chaotic maps $\Phi_{i},\;i=1,2,..$ with probability
$p_{i}\geq 0$ with $\sum p_{i}=1$, the corresponding random map
can be defined as
\begin{equation}
\Phi(x,p)=\Phi_{i}(x)\quad\quad\mbox{with probability $p_{i}$}
\end{equation}
 Here in this paper we try to constructs hierarchy of
random chaotic maps with an invariant measure by choosing the
ensemble of one and many parameters of chaotic maps of
trigonometric and elliptic types of references
\cite{jafar1,jafar2, jafar4} as follows.
\subsection{One-parameter random trigonometric maps }
The families of one-parameter trigonometric chaotic maps of the
interval $[0, 1]$ are defined as the ratio of hypergeometric
polynomials through the following equation \cite{jafar1}:
\begin{equation}
\Phi_{N}(x,\alpha)=\frac{\alpha^2\left(1+(-1)^N{
}_2F_1(-N,N,\frac{1}{2},x)\right)}
{(\alpha^2+1)+(\alpha^2-1)(-1)^N{ }_2F_1(-N,N,\frac{1}{2},x)}.
\end{equation}
where $N$ is an integer greater than one, obviously these map the
unit interval $[0, 1]$ into itself. Using the hierarchy of
families of one-parameter chaotic maps $(2.1)$, we can generate a
new hierarchy of random maps with an invariant measure, denote by
$\Phi(p_{i},N_{i}, \alpha,x)$ which can be written as:
\begin{equation}
\Phi\left(p_{i},N_{i},\alpha,x\right)=
\Phi_{N_{i}}\left(a_{N_{i}}(\alpha), x\right)\quad\quad \mbox{with
probability $p_{i}$},
\end{equation}
 where $ \sum^{m}_{i=1}{p_{i}}=1$ and the parameter $a_{N}(\alpha)$
 are defined as
\begin{equation}
a_{N}(\alpha)=\frac{\sum{^{[\frac{(N-1)}{2}]}_{k=0}
C^{N}_{2k+1}\left(\frac{\alpha}{1-\alpha}\right)^{-k}}}{\sum{^{[\frac{N}{2}]}_{k=0}
C^{N}_{2k}\left(\frac{\alpha}{1-\alpha}\right)^{-k}}},
\end{equation}
with the symbol $[\;]$ as greatest integer part, and
$C^n_m=\frac{n!}{m!(n-m)!}$.
\subsection{Many-parameter random trigonometric maps}
Even thought one can define many-parameters random trigonometric
chaotic maps with an invariant measure, but for simplicity we
restrict ourselves here in this paper to two-parameters ones
\cite{jafar2}. Random two-parameters trigonometric chaotic maps
are defined as:
\begin{equation}
\Phi(P_{ij}, \alpha_{i}, \alpha_{j}, N_{i}, N_{j}, x
)=\Phi_{N_{i}N_{j}}\left(\alpha_{i}, \alpha_{j},
x\right)\quad\quad \mbox{with probability $p_{ij}$},
\end{equation}
where $\sum_{i,j}P_{ij}=1$.
\begin{equation}
\Phi_{N_{i}N_{j}}(\alpha_{i}, \alpha_{j}, x)=
\Phi_{N_{i}}(\Phi_{N_{j}}(x,\alpha_{j}),\alpha_{i})
\end{equation}
with
\begin{equation}
\alpha^{-1}_{i}=\alpha_{j}\times
\frac{A_{N_{j}}(\alpha)}{B_{N_{j}}(\alpha)} \times
\frac{A_{N_{i}}\left(\frac{1}{\eta^{\alpha_{j}}_{N_{j}}(\alpha)}\right)}
{B_{N_{i}}\left(\frac{1}{\eta^{\alpha_{j}}_{N_{j}}\left(\alpha\right)}\right)}
\end{equation}
and
\begin{equation}
\eta^{\alpha_{j}}_{N_{j}}(\alpha)=\alpha_{j}\times(\frac{\alpha}{1-\alpha})
\times\left(\frac{A_{N_{j}}(\alpha)}{B_{N_{j}}(\alpha)}\right)^2,
\end{equation}
where the  polynomials $A(x)$, $B(x)$ are defined as:
 $$
 A(x)=\sum_{k=0}^{[ \frac{N}{2}]}C_{2k}^{N}{\left(\frac{x}{1-x}\right)}^{k},
$$
\begin{equation}
B(x)=\sum_{k=0}^{[
\frac{N-1}{2}]}C_{2k+1}^{N}{\left(\frac{x}{1-x}\right)}^{k}.
\end{equation}
\subsection{One-parameter random elliptic maps }
The families of one-parameter elliptic chaotic maps of $\bf{cn}$
\cite{wang} at the interval $[0,1]$ are defined as the ratio of
Jacobian elliptic functions of $\bf{cn}$ types in the following
form \cite{jafar4}:
\begin{equation}
 \Phi^{e}_{N}(x,\alpha)
=\frac{\alpha^2\left(cn\left(N
cn^{-1}(\sqrt{x})\right)\right)^{2}}{1+ (\alpha^2-1)\left(cn(N
cn^{-1}(\sqrt{x}))\right)^{2}}.
\end{equation}
 Obviously
these map the unit interval $[0,1]$ into itself. Now, with the
hierarchy of families of one-parameter elliptic chaotic maps
$(2.9)$, we can generate a new hierarchy of one-parameter random
elliptic maps with an invariant measure, denote by
$\Phi^{e}(p_{i},\alpha_{i},x)$ which can be written as:
\begin{equation}
\Phi^{e}(p_{i},N_{i},\alpha,x)=
\Phi^{e}_{N_{i}}(a_{N_{i}}(\alpha), x)\quad\quad \mbox{with
probability $p_{i}$},
\end{equation}
 where $ \sum^{m}_{i=1}{p_{i}}=1$ and $a_{N_{i}}(\alpha)$ is the
 same as given in (2.12).
\section{Invariant measure of random chaotic maps }
\setcounter{equation}{0} Characterizing invariant measure for a
given nonlinear dynamical systems is a fundamental problem which
connects dynamical theory to statistics and mechanics. A
well-known example is Ulam and von Neumann map which has an
ergodic measure $\mu=\frac{1}{\sqrt{x(1-x)}}$ \cite{Ulam}.\\ Let
us recall that for a deterministic map $\Phi(x)$, the invariant
probability measure $\mu(x)$ is the eigenfunction of the
Perron-Frobenius(PF) operator $\bf{L}$ related to maximum
eigenvalue 1 \cite{loreto, sinai}
\begin{equation}
L\mu(x)=\mu(x),
\end{equation}
where the operator $L$ is defined as:
\begin{equation}
Lf(x)=\int{\delta{\left(y-\Phi(x)\right)}f(y)}dy=\sum_{z=\Phi^{-1}(x)}\frac{f(z)}{\mid
\Phi^{'}(z)\mid }.
\end{equation}
In the case of random map, the average probability density can be
found by the straightforward generalization of $(3.1)$ as follows
\begin{equation}
\bar{L}\mu_{av}(x)=\mu_{av}(x),
\end{equation}
where
\begin{equation}
\bar{L}=\sum^{m}_{i=1}p_{i}L_i,
\end{equation}
where $L_i$ is the Perron-Frobenius operator associated with map
$\Phi_i(x)$.\\
 It should be mentioned that for trigonometric chaotic maps \cite{jafar1}, their
composition \cite{jafar2} and their coupling \cite{jafar3} the
eigenstate of PF operator $\bf{L}$ corresponding to largest
eigenvalues has already been obtained  in our previous papers.
Now,we choose the hierarchy of trigonometric chaotic maps $\Phi(
N_{i}, \alpha, x )$, as the  ensemble of chaotic maps. Then
$\Phi(p_{i}, N_{i}, \alpha, x )$-invariance of average density
$\mu_{av}(x, \alpha)$ implies that the average density should
satisfies the following formal Perron-Frobenius(PF) integral
equation
\begin{equation}
\mu_{av}(p, y, \alpha)=\sum^{m}_{i=1}p_{i}\int_{0}^{1}\delta
\left(y-\Phi_{N_{i}}(a_{N_{i}}(\alpha), x)\right) \mu_{i}(x,
\alpha)dx.
\end{equation}
Obviously above equation is the generalization of Equation $(3.2)$
for random trigonometric chaotic maps.
 As it is shown in\cite{jafar1}, each integral
 appearing on the right hand side of $(3.5)$ can be written as
\begin{equation}
\mu_{i}(y, \alpha)=\sum_{x_{ij}\varepsilon \Phi^{-1}_{N_{i}}(y,
a_{N_{i}}(\alpha) )}\mu_{i}{(x_{ij}, \alpha)}dx_{j}.
\end{equation}
Using the prescription of Reference \cite{jafar1} one can show
that $\mu_{i}(x, \alpha)$, the invariant measure associated with
trigonometric chaotic maps $\Phi( N_{i}, \alpha, x )$ has the
following form
\begin{equation}
 \mu_{i}{(x,\alpha)}= \mu{(x,\alpha)}=\frac{1}{\pi}\frac{\sqrt{\frac{\alpha}{1-\alpha}}}{\sqrt{x(1-x)}
\left(\frac{\alpha}{1-\alpha}+(1-\frac{\alpha}{1-\alpha})x\right)},
\end{equation}
that is, the invariant measure $\mu_{i}(x, \alpha)$ given in (3.7)
satisfies equation (3.6). Now, multiplying both side of equation
$(3.6)$ by $p_{i}$ and summing over $i$ we get
\begin{equation}
\mu_{av}(p,x, \alpha)=\sum^{m}_{i=1}p_{i}\mu_{i}(x,
\alpha)=\mu{(x,\alpha)}=
\sum^{m}_{i=1}p_{i}\times\sum_{x_{ij}\varepsilon \Phi^{-1}(p_{i},
N_{i}, \alpha, y )} \mu_{i}(x, \alpha)dx_{ij}
\end{equation}
Therefore, the density $\mu(x, \alpha)$ given in  $(3.7)$ is the
average invariant measure for ensemble of trigonometric chaotic
maps $\Phi_{i}(N_i,x)$ and it satisfies PF-equation $(3.5)$, hence
$\mu_{av}(x, \alpha)$ is the invariant or SRB measure \cite{sinai,
blank} of random  trigonometric chaotic maps maps given in $(2.4)$
define on the interval $[0, 1]$. Also, as the relation $(3.7)$
shows $\mu_{av}(x, \alpha)=\mu_{i}(x, \alpha)$, hence the average
invariant measure for random trigonometric chaotic maps is equal
to the invariant measure of each map of ensemble of chaotic
maps.\\ Also one can show that the average density $\mu_{av}(x,
\alpha)$ given in $(3.7)$ has the following asymptotic form of
delta function as $\alpha$ goes to zero and one, respectively,that
is, we have
\begin{equation}
\mu_{av}(x, \alpha)_{\alpha \rightarrow 0}=\delta(x),
\end{equation}
\begin{equation}
\mu_{av}(x, \alpha)_{\alpha \rightarrow 1}=\delta(x-1).
\end{equation}
where the first one corresponds to invariant measure associated
with fixed point $x=0$ and the latter one corresponds to the fixed
point $x=1$. It is straight forward to show that the random
trigonometric chaotic maps are well defined for $\alpha > 1 $,
where they have fix point ($x=1$)\cite{jafar1}, therefore, they
posses Dirac delta function  invariant measure for $\alpha
>1$, too.
\\ Similarly one can show that the average density of
two-parameters(many-parameters) random trigonometric chaotic maps
is the same as the average invariant measure  $\mu_{av}(x,
\alpha)$ given in $(3.7)$.\\ Finally  in the case of elliptic
random maps, as it is shown in reference \cite{jafar4}, for small
values of elliptic parameter $\bf{k}$, elliptic chaotic maps are
topologically conjugated with trigonometric chaotic maps. Hence,
for small $\bf{k}$ the average invariant measure for one-parameter
random elliptic chaotic maps of $\bf{cn}$ type, is also the same
as the average invariant measure  $\mu_{av}(x, \alpha)$ given in
$(3.7)$.
\section{Kolmogrov-Sinai entropy of random chaotic maps}
\setcounter{equation}{0}
 KS-entropy or metric entropy \cite{sinai} measures how
chaotic a dynamical system is and it is proportional to the rate
at which information about the state of dynamical system is lost
in the course of time or iteration. Therefore, it can also be
defined as the average rate of loss of information for a discrete
measurable dynamical system $(\Phi(x, p),\mu_{av})$. By
introducing a partition $\alpha={A_c} (n_1,.....n_{\gamma})$ of
the interval $[0,1]$ into individual laps $A_i$, one can define
the usual entropy associated with the partition by:
$$H(\mu_{av},\gamma)=-\sum^{n(\gamma)}_{i=1}m(A_c)\ln{m(A_c)},$$
where $m(A_c)=\int{_{n\in{A_i}}\mu_{av}(x, \alpha)dx}$ is the
invariant measure of $ A_i$. Defining a n-th refining $\gamma(n)$
of  $ \gamma$: $$\gamma^{n}=\bigcup^{n-1}_{k=0}(\Phi(x,
p))^{-(k)}(\gamma)$$ then an entropy per unit step of refining is
defined by : $$h(\mu_{av},\Phi(x, p),\gamma)=
\lim{_{n\rightarrow{\infty}}}\left(\frac{1}{n}H(\mu_{av},\gamma)\right),$$
now, if the size of individual laps of $\gamma(N)$ tends to zero
as n increases, the above entropy reduces to well known as
KS-entropy, that is: $$h(\mu_{av},\Phi(x, p))= h(\mu_{av},\Phi(x,
p),\gamma).$$
 KS-entropy is actually a quantitative measure of the rate of information
lost with the refining and it can be written as \cite{loreto}:
\begin{equation}
h(\mu_{av},\Phi(x,p)=\sum^{m}_{i=1}{p_{i}}
\int{\mu_{av}(x)}\ln{\mid\frac{d\Phi_{i}(x)}{dx}\mid}dx,
\end{equation}
which is also a statistical mechanical expression of the Lyapunov
Characteristic exponent, that is: mean divergence rate of two
nearby orbits. The measurable random dynamical system $(\Phi(p,
 x ),\mu_{av})$ is chaotic for $h(\mu_{av},\Phi(x,p))>0$ and predictive for
$h(\mu_{av},\Phi(x,p))=0$. Using the fact that the invariant
measure for these random chaotic maps is equal to invariant
measure of each map of ensemble of chaotic maps, one can show that
KS-entropy of these random chaotic maps is the average KS-entropy
of chaotic maps of ensemble, that is we have:
\begin{equation}
h(\mu_{av},\Phi(x, p))=\sum^{m}_{i=1}{p_{i}}\times
h(\mu_{i},\Phi_{i}(x)).
\end{equation}
\subsection{KS-entropy of one-parameter random trigonometric chaotic maps }
With a prescription similar to the prescription of Reference
\cite{jafar1}, one can calculate KS-entropy of hierarchy of
trigonometric chaotic maps $\Phi(a_{N_{i}}(\alpha), x)$, where we
quote only the result below
\begin{equation}
h\left(\mu_{i},\Phi( N_{i}, \alpha, x)\right)=
 \ln\left(\frac{N_{i}(\frac{1}{1-\alpha}+
 2\sqrt{\frac{\alpha}{1-\alpha}})^{N_{i}-1}}{(\sum_{k=0}^{[ \frac{N_{i}}{2}]}
 C_{2k}^{N_{i}}(\frac{\alpha}{1-\alpha})^{k})(\sum_{k=0}^{[
 \frac{N_{i}-1}{2}]}C_{2k+1}^{N_{i}}(\frac{\alpha}{1-\alpha})^{k})}\right).
\end{equation} Therefore, substituting for  KS-entropy of
one-parameter trigonometric chaotic map in  Equation $(4.1)$, we
get the following expression for KS-entropy of one-parameter
random trigonometric chaotic maps $h\left(\mu_{av},\Phi(p_{i},
N_{i}, \alpha, x)\right)$
\begin{equation}
h\left(\mu_{av},\Phi(p_{i}, N_{i}, \alpha,
x)\right)=\sum^{m}_{i=1}p_{i}
 \ln\left(\frac{N_{i}(\frac{1}{1-\alpha}+
 2\sqrt{\frac{\alpha}{1-\alpha}})^{N_{i}-1}}{(\sum_{k=0}^{[ \frac{N_{i}}{2}]}
 C_{2k}^{N_{i}}(\frac{\alpha}{1-\alpha})^{k})(\sum_{k=0}^{[
 \frac{N_{i}-1}{2}]}C_{2k+1}^{N_{i}}(\frac{\alpha}{1-\alpha})^{k})}\right).
\end{equation}
Using the asymptotic Dirac delta function form of the average
density $\mu_{av}(x, \alpha)$ for limiting values of
$\alpha=o,\;\mbox{and}\;1$  given in (3.9) and (3.10),
respectively, one can show that, KS-entropy of one-parameter
random trigonometric chaotic maps takes the following form:
\begin{equation}
h\left(\mu_{av},\Phi(p_{i}, N_{i}, \alpha, x
)\right)=\sum^{m}_{i=1}p_{i} \ln{\mid \frac{d \Phi_{N_{i}}\left(
a_{N_{i}}(\alpha), x
\right)}{dx}|_{x=0}\mid}=\sum^{m}_{i=1}p_{i}\ln{\mid
\frac{N_{i}}{a^{2}_{N_{i}}(\alpha)}\mid}=0.
\end{equation}
as $\alpha \rightarrow 0$, and
\begin{equation}
 h\left(\mu_{av},\Phi(p_{i}, N_{i},
\alpha, x )\right)=\sum^{m}_{i=1}p_{i} \ln{\mid \frac{d
\Phi_{N_{i}}( a_{N_{i}}(\alpha), x
)}{dx}|_{x=1}\mid}=\sum^{m}_{i=1}p_{i}\ln{\mid
N_{i}a^{2}_{N_{i}}(\alpha) \mid}.
\end{equation}
as $\alpha \rightarrow 1$ and for $\alpha >1$, respectively.\\ It
is straight forward to see that each summed on the right hand side
of $(4.4)$ has the asymptotic form $(1-\alpha)^{\frac{1}{2}}$ as
$\alpha \rightarrow 1_{-}$. Thus $h(\mu_{av},\Phi(p_{i}, N_{i},
\alpha, x ))$ has the following asymptotic form as $(\alpha
\rightarrow 1_{-})$.
\begin{equation}
\left\{ \begin{array}{l} h\left(\mu_{av},\Phi\left(p_{i}, N_{i},
\alpha\rightarrow 1_{-},x\right)\right)\sim
(1-\alpha)^{\frac{1}{2}},\\\\ h\left(\mu_{av},\Phi\left(p_{i},
N_{i}, \alpha\rightarrow 0_{+},x\right)\right)\sim
(\alpha)^{\frac{1}{2}} ,
\end{array}\right.
\end{equation}
The above asymptotic form indicates that the maps $\Phi(p_{i},
N_{i}, \alpha, x )$ belong to the same universality class which
are different from the universality class of pitch fork
bifurcating maps but their asymptotic behavior is similar to class
of intermittent maps\cite{pomeau}, even though intermittency can
not occur in these maps for any values of parameter
$a_{N}(\alpha)$, since the maps $\Phi(p_{i}, N_{i}, \alpha, x )$
and their n-composition $\Phi^{(n)}$ do not have minimum values
other than zero and maximum values other than one in the interval
$[0,1].$
\subsection{KS-entropy of many-parameter random trigonometric chaotic maps }
 Similarly, one can calculate KS-entropy of
many-parameter random chaotic maps with the prescription reference
\cite{jafar2}:
 $$h\left(\mu_{av},\Phi(p_{ij}, N_{i}, N_{j},
\alpha_{i},\alpha_{j}, x )\right)=$$
\begin{equation}
\sum_{i,j}p_{ij}\ln
\left(\frac{N_iN_j(1+\sqrt{\frac{1-\alpha}{\alpha}})^{2(N_{j}-1)}
(1+\sqrt{\eta_{N_{j}}^{\alpha_{j}}(\alpha)})^
{2(N_{i}-1)}}{A_{N_j}(\alpha)B_{N_j}(\alpha)
A_{N_{i}}(\eta_{N_{j}}^{\alpha_{j}}(\alpha))
B_{N_{i}}(\eta_{N_{j}}^{\alpha_{j}}(\alpha))}\right).
\end{equation}
With respect to the one-parameter random trigonometric chaotic
maps, the numerical and theoretical calculations predict different
asymptotic behavior for many-parameter random trigonometric
chaotic maps, as example of asymptotic of the composed maps
($\phi_{2,3}(\alpha_{1},\alpha_{2},x)$ and
$\phi_{3,2}({\alpha}_{1},\acute{\alpha}_{2},x)$), the KS-entropy
$h(\mu_{av},\Phi)$ is presented below: $$
h(\mu_{av},\Phi)=p\times\ln{\left(\frac{3\left((1-\alpha)+\sqrt{\alpha(1-\alpha)}\right)^4
\left((2\alpha+1)
(1-\alpha)+\alpha_{2}^{2}(3-2\alpha)\sqrt{\alpha(1-\alpha)}\right)^2}{(1-\alpha)^3(1+2\alpha)
(3-2\alpha)((1-\alpha)(1+2\alpha)^2+\alpha_{2}^2\alpha(3-2\alpha)^2)}\right)}+
$$
\begin{equation}
(1-p)\times\ln{\left(\frac{3((1-\alpha)+\sqrt{\alpha(1-\alpha)})^4((2\alpha+1)
(1-\alpha)+\acute{\alpha}_{2}^{2}(3-2\alpha)\sqrt{\alpha(1-\alpha)})^2}{(1-\alpha)^3(1+2\alpha)
(3-2\alpha)((1-\alpha)(1+2\alpha)^2+\acute{\alpha_{2}}^2\alpha(3-2\alpha)^2)}\right)}
\end{equation}
 with the following relation among the parameters
 $$\alpha_{2}^{-1}=\frac{\alpha_{1}(1-\alpha)(1+2\alpha)^3}{2(3-2\alpha)((1-\alpha)
 (1+2\alpha)-\alpha_{1}\alpha(3-2\alpha)^2)}$$
  $$\acute{\alpha}_{2}^{-1}=\frac{{\alpha}_{1}(1-\alpha)(1+2\alpha)^3}
  {2(3-2\alpha)((1-\alpha)(1+2\alpha)-{\alpha_{1}}\alpha(3-2\alpha)^2)}$$
 which is obtained from the relation $(2.8)$. Now choosing
$\alpha_{2}=\acute{\alpha_{2}}$ and
$\alpha=\frac{\alpha_2^{\nu}}{1+\alpha_2^{\nu}} \>, 0<\nu<2$, the
 entropy given by $(4.9)$ reads:
$$h=p\times\ln{\left(\frac{3(1+\sqrt{\alpha_{2}^{\nu}})^4(1+3\alpha_{2}^{\nu})(1+\alpha_{2}^{\nu+2}
\sqrt{\alpha_{2}^{\nu}})}{(3+\alpha_{2}^{\nu})(1+3\alpha_{2}^{\nu})((3\alpha_{2}^{\nu}+1)+
\alpha_{2}^{\nu+2}(3+\alpha_{2}^{\nu})^2)}\right)}+$$
\begin{equation}
(1-p)\times\ln{\left(\frac{3(1+\sqrt{\alpha_{2}^{\nu}})^4(1+3\alpha_{2}^{\nu})(1+\alpha_{2}^{\nu+2}
\sqrt{\alpha_{2}^{\nu}})}{(3+\alpha_{2}^{\nu})(1+3\alpha_{2}^{\nu})((3\alpha_{2}^{\nu}+1)+
\alpha_{2}^{\nu+2}(3+\alpha_{2}^{\nu})^2)}\right)}
\end{equation}
 which has the following asymptotic behavior
 $$ \left\{ \begin{array}{l} h(\mu_{av},\Phi)\sim\alpha_2^{\frac{\nu}{2}}\quad
\quad\mbox{as}\quad \alpha_2\longrightarrow
0(\alpha\longrightarrow 0),
\\\\ h(\mu_{av},\Phi)\sim(\frac{1}{\alpha_2})^
{\frac{\nu}{2}}\quad\quad\mbox{as}\quad
\alpha_2\longrightarrow\infty(\alpha \longrightarrow 1).
\end{array}\right.$$

 The above asymptotic behaviours indicate that for an  arbitrary value
of $0<\nu<2$ the maps $\Phi(p_{ij}, 2, 3, \alpha_{1},\alpha_{2}, x
)$ belong to the universality class which is different from the
universality class of one-parameter trigonometric chaotic maps of
$\Phi(p_{i}, N_{i},\alpha, x)$ $(2.2)$ or the universality class
of pitch fork bifurcating maps.
\subsection{KS-entropy of one-parameter random elliptic chaotic maps }
 For random one-parameter elliptic
chaotic maps of $\bf{cn}$ type, KS-entropy, for small values of
elliptic parameter by considering random elliptic chaotic maps are
topologically conjugated with random trigonometric chaotic maps
\cite{jafar4} would be equal to KS-entropy of one-parameter random
trigonometric chaotic maps, is represented with $(4.3)$.
\section{Lyapunov exponent of random chaotic maps}
 The Lyapunov exponent $\lambda$ provides the simplest
information about chaoticty and can be computed considering the
separation of two nearby trajectories evolving in the same
realization of the random process, and for random chaotic maps
given in $(2.3,5,11)$, it can be defined as  \cite{Dorf}
\begin{equation}
\lambda(x_{0})=lim_{n\rightarrow\infty}\frac{1}{n}\sum_{k=0}^{n-1}\ln\mid
\frac{d\Phi(x ,p)}{dx}\mid,
\end{equation}
where $x_{k}=\overbrace{\Phi_{N} \circ \Phi_{N}\circ ....\circ
\Phi_{N}^{k}(x_{0})}$. It is obvious that its negative values,
indicate that the system is in  fix point (attractor)regime, while
its positive values indicate  that the  system is  measurable( the
Invariant measure given in  $(3.9, 12)$) \cite{Dorf}. Also, the
lyapunov number is independent of initial point, provided that the
motion inside the invariant manifold is ergodic, thus
$\lambda{(x_{0})}$ characterizes the invariant manifold of random
map as a whole. For values of parameter $\alpha$, such the map
$\Phi(x, p)$ be measurable, Birkhof ergodic \cite{keller} theorem
implies the equality of KS-entropy and lyapunov number, i.e.,
 $$h(\mu_{av}, \Phi(x, p))=\lambda(x_{0})$$
 Also comparing KS-entropy of these maps
by their Lyapunov exponent confirms this prediction (see figures
1a,1b,2a and 2b). In chaotic region, random maps are ergodic as
Birkhof ergodic theorem predicts. In non-chaotic region of the
parameter, lyapunov characteristic exponent is negative definite,
since in this region, we have only single period fixed points
without bifurcation.
\section {Conclusion}
In this paper we have discussed the dynamical characterization of
systems whose evolution is described by random maps. We have
studied the application of Perron-Frobenius operator to the
analysis of the dynamical behaviour of random dynamical systems in
order to derive the invariant measure of the system. Again this
interesting property is due to the existence of invariant measure
for a region of the parameters space of these maps.

 \newpage
{\bf Figure Captions}
\\ Fig.1-a. Shows the variation of KS-entropy of one-parameter random trigonometric chaotic map
 for ensemble of ($\Phi_{2}(x,a_{2}(\alpha))$ and
$\Phi_{3}(x,a_{3}(\alpha))$) in term of the parameter $\alpha$ and
p
\\ Fig.1-b. Shows the variation of Lyapunov characteristic exponent of one-parameter random trigonometric chaotic map
 for ensemble of ($\Phi_{2}(x,a_{2}(\alpha))$ and
$\Phi^{e}_{3}(x,a_{3}(\alpha))$) in term of the parameter $\alpha$
and p
\\ Fig.2-a. Shows the variation of KS-entropy of one-parameter random elliptic chaotic maps
 for ensemble of ($\Phi^{e}_{2}(x,a_{2}(
\alpha))$ and $\Phi^{e}_{3}(x, a_{3}(\alpha))$)  in term of the
parameter $\alpha$ and p
\\ Fig.2-b. Shows the variation of  Lyapunov characteristic exponent of one-parameter random
elliptic chaotic maps for ensemble of ($\Phi^{e}_{2}(x,a_{2}(
\alpha))$ and $\Phi^{e}_{3}(x, a_{3}(\alpha))$)  in term of the
parameter $\alpha$ and p d

\begin{thebibliography}{99}
\small{
\bibitem{Ulam}{\sc S. M. Ulam and J. Von Neumann, }{\em Bull. Am. Math. Soc. 53, (1974) 1120-1127.}
\bibitem{Adler}{\sc R. L. Adler and T. J. Rivlin, }{\em Proc. Am. Math. Soc. 15, (1964) 794-799.}
\bibitem{Kasta}{\sc S. Katsura and W. Fukuda, }{\em Physica 130 A, (1964) 597-601.}
\bibitem{Gyo}{\sc G. Gyorgyi and P. Szepfalusy, }{\em Phys. B 55, (1984) 179-184.}
\bibitem{dorf1}{\sc R. Klages and J. R. Dorfman, }{\em Phys. Rev lett 74, (1995) 387-390.}
\bibitem{umeno1}{K. Umeno, }{\em Phys. Rev. E 55, (1997) 2644-2647.}
\bibitem{jafar3}{\sc M. A. Jafarizadeh and S. Behnia, }{\em Physica D 159, (2001) 1-21.}
\bibitem{jafar1}{\sc M. A. Jafarizadeh, S. Behnia, S. Khorram and H. Naghshara,}
 {\em Journal of statistical physics, Vol. 104, (2001) 1013-1028.}
\bibitem{jafar2}{\sc  M. A. Jafarizadeh and S. Behnia, }{\em Journal of nonlinear physics, Vol.
9, (2002) 26-41.}
\bibitem{jafar4}{\sc M. A. Jafarizadeh and S. Behnia, }{\em  Hierarchy of one and many parameters
families of elliptic chaotic maps of $\bf{cn}$ and $\bf{sn}$
types, nlin. CD/0208024 (Submitted to Physics Letter A). }
\bibitem{coste}{\sc J. Cosete and M. Henon, }{\em In disorderd systems and biological organization,
 Springer-Verlag, (1988). }
\bibitem{Derrida}{\sc B. Derrida and H. Flyvbjerg, }{\em J. Physique 48, (1987) 971-978.}
\bibitem{ott}{\sc L. Yu, E. Ott and Q. Chen, }{\em Physical review letters, Vol.65,(1990) 2935-2938. }
\bibitem{wang}{\sc Z. X. Wang and  D. R. Guo, }{ \em Special
Functions,
 World Scientific Publishing, (1989).}
 \bibitem{loreto}{\sc V. Loreto, G. Paladin, and A. Vulpiani, }{\em Review E 53, (1996) 2085-2098. }
\bibitem{sinai}{\sc I. P. Cornfeld, S. V. Fomin and Ya. G. Sinai,}{
\em Ergodic Theory. Springer-Verlag, (1982).}
\bibitem{blank}{\sc M. Blank, }{\em Moscow mathematical journal Vol. 1, (2001) 315-344.}
\bibitem{pomeau}{\sc Pomeau Y, and Manneville P. }{\em Communications in Mathematical Physics, (1980) 74-86.}
\bibitem{Dorf}{\sc J. R. Dorfman, }{\em An Introduction to chaos in
nonequilibrium statistical mechanics, Cambridge (1999).}
\bibitem{keller}{\sc G. keller, }{\em Equilibrium state in ergodic theory,
Cambrige university press, (1998).}
 \bibitem{dev}{\sc R. L. Devancy, }{ \em An Introduction to Chaotic
Dynamical Systems, Addison Wesley, (1982).}}
\end{thebibliography}
\end{document}